\documentclass[aps,prd,twocolumn,groupedaddress]{revtex4}
  \usepackage{subfigure} 
\usepackage{graphicx}
  \usepackage{subfigure} 

\def\jnl@style#1{{\rmfamily#1}}%
\def\jref@jnl#1{{\jnl@style#1}}%

\newcommand\aj{\jref@jnl{AJ}}%
\newcommand\araa{\jref@jnl{ARA\&A}}%
\newcommand\apjl{\jref@jnl{ApJ}}%
\newcommand\apjs{\jref@jnl{ApJS}}%
\newcommand\apss{\jref@jnl{Ap\&SS}}%
\newcommand\aap{\jref@jnl{A\&A}}%
\newcommand\aapr{\jref@jnl{A\&A~Rev.}}%
\newcommand\aaps{\jref@jnl{A\&AS}}%
\newcommand\azh{\jref@jnl{AZh}}%
\newcommand\baas{\jref@jnl{BAAS}}%
\newcommand\jrasc{\jref@jnl{JRASC}}%
\newcommand\memras{\jref@jnl{MmRAS}}%
\newcommand\mnras{\jref@jnl{MNRAS}}%
\newcommand\pasp{\jref@jnl{PASP}}%
\newcommand\pasj{\jref@jnl{PASJ}}%
\newcommand\qjras{\jref@jnl{QJRAS}}%
\newcommand\skytel{\jref@jnl{S\&T}}%
\newcommand\solphys{\jref@jnl{Sol.~Phys.}}%
\newcommand\sovast{\jref@jnl{Soviet~Ast.}}%
\newcommand\ssr{\jref@jnl{Space~Sci.~Rev.}}%
\newcommand\zap{\jref@jnl{ZAp}}%
\newcommand\iaucirc{\jref@jnl{IAU~Circ.}}%
\newcommand\aplett{\jref@jnl{Astrophys.~Lett.}}%
\newcommand\apspr{\jref@jnl{Astrophys.~Space~Phys.~Res.}}%
\newcommand\bain{\jref@jnl{Bull.~Astron.~Inst.~Netherlands}}%
\newcommand\fcp{\jref@jnl{Fund.~Cosmic~Phys.}}%
\newcommand\gca{\jref@jnl{Geochim.~Cosmochim.~Acta}}%
\newcommand\grl{\jref@jnl{Geophys.~Res.~Lett.}}%
\newcommand\jgr{\jref@jnl{J.~Geophys.~Res.}}%
\newcommand\jqsrt{\jref@jnl{J.~Quant.~Spec.~Radiat.~Transf.}}%
\newcommand\memsai{\jref@jnl{Mem.~Soc.~Astron.~Italiana}}%
\newcommand\nphysa{\jref@jnl{Nucl.~Phys.~A}}%
\newcommand\physrep{\jref@jnl{Phys.~Rep.}}%
\newcommand\physscr{\jref@jnl{Phys.~Scr}}%
\newcommand\planss{\jref@jnl{Planet.~Space~Sci.}}%
\newcommand\procspie{\jref@jnl{Proc.~SPIE}}%
\newcommand{\sm}[1]{\mbox{{\scriptsize #1}}}
\newcommand{\simle} {\,{}^<_{\sim}\,}

\newcommand{\HI}{$\mathrm{H}$\ }
\newcommand{\HII}{$\mathrm{H}^+$\ }

\newcommand{\HeI}{$\mathrm{He}$\ }
\newcommand{\HeII}{$\mathrm{He}^+$\ }

\newcommand{\HId}{$\mathrm{H}$}

\newcommand{\fHI}{\mathrm{H}}
\newcommand{\fHII}{\mathrm{H}^+}

\newcommand{\fHeI}{\mathrm{He}}

\newcommand{\fe}{\mathrm{e}}
\newcommand{\fpp}{\mathrm{p}}

\begin{document}


\title{Reionization - A probe for the stellar population and the physics of the early universe.}


\author{Dominik R. G. Schleicher, Robi Banerjee, Ralf S. Klessen}
\email{dschleic@ita.uni-heidelberg.de}
\affiliation{Institute of Theoretical Astrophysics / ZAH,\\ Albert-Ueberle-Str. 2,\\ D-69120 Heidelberg,\\ Germany}


\date{\today}

\begin{abstract}
We calculate the reionization history for different models of the stellar population and 
explore the effects of primordial magnetic fields, dark matter decay and dark matter annihilation
on reionization. We find that stellar populations based on a Scalo-type initial mass function for Population II stars can be ruled out as sole sources for reionization, unless star formation efficiencies of more than $10\%$ or very high photon escape fractions from the parental halo are adopted. 
When considering primordial magnetic fields, we find that the additional heat injection from ambipolar diffusion and decaying MHD turbulence has significant impact on the thermal evolution and the ionization history of the post-recombination universe and on structure formation. The magnetic Jeans mass changes the typical mass scale of the star forming halos,  and depending on the adopted stellar model we derive upper limits to the magnetic field strength between  $0.7$ and $5\,$nG (comoving).
For dark matter annihilation, we find an upper limit to the thermally averaged mass-weighted cross section of $\langle\sigma v \rangle/m_{DM}\le10^{-33}\ \mathrm{cm}^3\mathrm{/s/eV}$. For dark matter decay, our calculations yield a lower limit to the lifetime of dark matter particles of $\tau\ge3\times10^{23}\,$s. These limits are 
in agreement with constraints from recombination and provide an independent confirmation at a 
much later epoch. 
\end{abstract}

\maketitle

\section{Introduction}
Over the last decades, our understanding of cosmological reionization has improved considerably. Observations of high-redshift quasars clearly indicate that reionization must end around $z\sim6$ \citep{Becker}, and the WMAP 5-year-measurement finds a reionization optical depth of $\tau=0.087\pm0.017$ \citep{Komatsu, WMAPAngular}, yielding clear evidence that reionization started significantly earlier and is thus a continuous process. Reionization through miniquasars would produce a soft-X-ray background that is significantly higher than the observed background, and can thus be ruled out \citep{Dijkstra, Salvaterra}. Simulations of stellar reionization further show that the process is highly inhomogeneous and based on the growth and merging of ionized bubbles around the first stellar sources \citep{Gnedin, Ciardi2, Kohler}. Understanding the essential physical processes, it is possible to build semi-analytic models that describe stellar reionization, which can be tested over a wide parameter space \citep{Shapiro, Haiman, Barkana, Loeb, Choudhury, Schneider}. Additional constraints on the cosmic star formation rate are now available from gamma-ray burst observations \citep{Yuksel}. 
The reionization framework thus provides an increasingly
reliable test for the stellar population during
reionization and can be used to constrain global physical 
conditions in the early universe.\\ \\

Regarding the stellar population, the works \cite{Abel} and \cite{Bromm} suggest that the first stars were top-heavy with a peak in the IMF at around $100\ M_\odot$. On the contrary, \cite{Clark} and \cite{Omukai} indicate that gas especially in more massive systems can fragment because of dips in the equation of state, which may lead to the formation of a stellar cluster. It was shown that cooling in previously ionized gas is enhanced and leads to typical stellar masses of $\sim10\ M_\odot$ \cite{Yoshida1, Yoshida2}. It was further suggested that the presence of weak magnetic fields is sufficient to lead to a more present-day like mode of star formation, resulting in considerably lower masses \citep{Silk}. Recently, a new phase of stellar evolution was suggested in which the stars would be powered by dark matter annihilation instead of nuclear fusion \cite{Spolyar}. Various follow-up works have explored such a scenarios in more detail. Studies by \citet{Iocco, Freese} explored the main-sequence and pre-main-sequence phase of dark stars, and other works calculated the effect of dark matter capture by off-scattering from baryons \citep{IoccoBressan,Yoon,FreeseSpolyar,Taoso}. While many predictions are still model-dependent, it has often been suggested that such stars may have typical masses of $800\ M_\odot$, giving rise to a very bright main-sequence phase, or may have much longer lifetimes due to modifications in the stellar evolution. In the end, all these suggestions must face the constraint that the stellar population must be able to provide the correct reionization optical depth. \\ \\

Complications may arise through the presence of additional physics that are often not considered in standard reionization calculations and simulations on the first stars. Such possibilities include the presence of primordial magnetic fields, dark matter decay and dark matter annihilation. Magnetic fields have been observed on all scales in the universe, 
and recently, it has been demonstrated that they were present already in high-redshift galaxies \citep{Bernet}. They are
found in the interstellar gas as well as in the intergalactic
medium 
\citep[][]{Beck96, Beck01, Carilli02}, but their origin
is still unclear. There is a viable possibility
that these fields have a primordial origin \citep{Grasso01,
Widrow02}. 
So far, the most stringent constraints on the strength of
these putative fields come from the measurements of the cosmic
microwave background radiation (CMBR) and from big-bang
nucleosynthesis (BBN) calculations.  A homogeneous magnetic field
would produce temperature anisotropies in the CMBR~\citep{Zeldovich83}
whose maximal amplitudes are limited by the COBE satellite
measurements which in turn limit the field strength, as measured
today, to $B_0 \simle 3.5\times 10^{-9} \,
\rm{G}$~\citep{Barrow97}. The presence of any primordial field would
also alter the CMBR power spectrum by changing the characteristic
velocities. With the sensitivity of the PLANCK satellite \footnote{http://sci.esa.int/science-e/www/area/index.cfm?fareaid=17} one should be
able to detect fields with present day strength of $B_0 > 5\times
10^{-8} \, \rm{G}$~\citep{Adams96}.  So far, measurements by the WMAP
satellite \footnote{http://map.gsfc.nasa.gov/} are compatible with the absence of primordial magnetic
fields.

Strong magnetic fields in the early universe can also change the
abundance of relic $^4$He and other light elements during the big bang
nucleosynthesis~\citep{Matese69, Greenstein69}. To
comply with observational limits on light element abundances these
primordial fields must not exceed $10^{12} \, \rm{G}$ at the time when
the universe was $T = 5\times 10^9 \, \rm{K}$ which
corresponds to a present day field $B_0 \simle 3\times 10^{-7} \,
\rm{G}$\cite{Greenstein69}. 

Effects of primordial magnetic fields have already been considered in \cite{Kim}, finding that density perturbations can be enhanced by the Lorentz force from tangled magnetic fields. Both the evolution of perturbations in the presence of magnetic fields as well as their effect on the thermodynamics via ambipolar diffusion heating and decaying MHD turbulence was considered in \cite{Sethi05}. Recent calculations of \cite{Tashiro} show that the enhancement of structure due to magnetic fields is pronounced at about $5\times10^6\ M_\odot$, but becomes less effective on larger mass scales and appears as a subdominant contribution on the scale of the magnetic Jeans mass. Consequences for $21$ cm observations have been explored as well \cite{Tashiro21,Schleicher}. In this work, we examine the consequences for reionization in more detail and calculate the backreaction on structure formation according to the work of 
\citep{Gnedin}. The WMAP 5 year data \citep{WMAPAngular} have measured
the Thomson scattering optical depth from reionization and allow to
constrain different reionization scenarios in the early universe. 

The nature of dark matter is still unclear, and consequences of various particle physics like massive neutrinos or axion decay have been explored early \citep[e. g.][]{Doroshkevich,Berezhiani, Berezhiani2}. Observational progress allowed to refine these studies and to explore such scenarios in the framework of $\Lambda$CDM cosmology \citep{Chen, Hansen, Avelino, Kasuya, Pierpaoli, Bean, Padmanabhan}. Recently, the possibility was discussed to detect such effects using future $21$ cm telescopes\cite{FurlanettoDM}, and it was shown that the fraction of the energy absorbed into the IGM can be calculated in detail for specific models of dark matter decay and annihilation \cite{Ripamonti}. Indeed, secondary ionization through the decay / annihilation products can provide a way to ionize the IGM which is independent of the stellar contribution \citep{Mapelli}, whereas the additional heat input increases the Jeans mass and delays the formation of the first structures.

In this work, we show how reionization constrains the properties of the stellar population as well as some additional heat sources like primordial magnetic fields, dark matter annihilation and decay. In \S\ref{reionization}, we present our model for stellar reionization, which considers the IGM as a two-phase-medium consisting of ionized bubbles and overall neutral gas. Based on the thermal evolution, we self-consistently determine the minimal mass scale of halos which can collapse. In \S \ref{bfields}, we explain our treatment of primordial magnetic fields and show how they modify the thermal evolution. The treatment of dark matter annihilation and decay, as well as some implications for dark stars, are discussed in \S \ref{darkness}. The optical depth for different models is given in \S \ref{cmbfast}. Further discussion and outlook is given in \S \ref{outlook}.

\section{Reionization in the early universe}\label{reionization}
The thermal and ionization history of the IGM between recombination
and the end of reionization is determined by a number of different processes. At
redshifts $z>300$, Compton scattering of CMB photons couples the
gas temperature $T$ to the CMB temperature $T_{\sm{rad}}$. At lower redshifts,
this coupling is less efficient and the gas decouples from the CMB due to adiabatic
expansion. This standard scenario can be altered if additional energy injection mechanisms are present.
Additional heat input can lead to an earlier redshift of decoupling, whereas an increase in the ionized 
fraction, for instance due to secondary ionization from the decay or annihilation products of dark matter, tends to make 
Compton cooling more efficient. Once structure formation sets in, the thermal and ionization history is further influenced
from X-rays produced in star forming regions and UV photons that
escape from the first galaxies. We modified the RECFAST
code~\citep{SeagerFast} and included all these feedback processes as described below, modeling the IGM as a two-phase medium of ionized and partially-ionized gas. In this picture, the fully-ionized gas refers to the gas in the HII regions of the first luminous sources, while the partially ionized gas describes gas that is not yet affected by UV feedback. 

\subsection{The RECFAST code}
In the absence of additional energy injection mechanisms, recombination in the early universe and freeze-out of electrons was calculated with unprecedent accuracy, following the detailed level populations of hundreds of energy levels for \HId, \HeI and \HeII and self-consistently calculating the radiation field\cite{Seager}. They developed the RECFAST code \footnote{http://www.astro.ubc.ca/people/scott/recfast.html}, a simplified version of the multi-level calculations, based on an effective three-level model for the hydrogen atom. RECFAST is capable of fully reproducing the results of the more detailed calculation \citep{SeagerFast}. Both the detailed calculation and the RECFAST code were recently updated and improved in \cite{Wong}. For the partially-ionized gas, we modify the equations describing the thermal and ionization history in the following way: The equation for the temperature evolution is given as
\begin{eqnarray}
\frac{dT}{dz}&=&\frac{8\sigma_T a_R T_{\sm{rad}}^4}{3H(z)(1+z)m_e c}\frac{x_e\,(T-T_{\sm{rad}})}{1+f_{\fHeI}+x_e}\nonumber\\
&+&\frac{2T}{1+z}-\frac{2(L_{\sm{heat}}-L_{\sm{cool}})}{3nk_B H(z)(1+z)},\label{temp}
\end{eqnarray}
where $L_{\sm{heat}}$ is the new heating term (see \S \ref{ambi}, \S \ref{decay} and \S \ref{stellar}), $L_{\sm{cool}}$ the new cooling term including Lyman $\alpha$ cooling, bremsstrahlung cooling and recombination cooling using cooling functions of \cite{Anninos}, $\sigma_T$ is the Thomson scattering cross section, $a_R$ the Stefan-Boltzmann radiation constant, $m_e$ the electron mass, $c$ the speed of light, $k_B$ Boltzmann's constant, $n$ the total number density, $x_e=n_\fe/n_\fHI$ the electron fraction per hydrogen atom, $H(z)$ is the Hubble factor and $f_{\fHeI}$ is the number ratio of \HeI and \HI nuclei, which can be obtained as $f_{\fHeI}=Y_p/4(1-Y_p)$ from the mass fraction $Y_p$ of \HeI with respect to the total baryonic mass.
The evolution of the ionized fraction of hydrogen, $x_\fpp$, is given as
\begin{eqnarray}
\frac{dx_\fpp}{dz}&=&\frac{[C(z)x_\fe x_\fpp n_\fHI \alpha_\fHI -\beta_\fHI (1-x_\fpp)e^{-h_p\nu_{\fHI, 2s}/kT}] }{H(z)(1+z)[1+K_\fHI(\Lambda_\fHI+\beta_\fHI)n_\fHI(1-x_\fpp)]}\\
&\times&[1+K_\fHI\Lambda_\fHI n_\fHI(1-x_\fpp)]-\frac{k_{\sm{ion}n_H x_\fpp}}{H(z)(1+z)}-f_{\sm{ion}}\nonumber.\label{ion}
\end{eqnarray}
In this equation, $n_\fHI$ is the number density of hydrogen atoms and ions, $h_p$ Planck's constant, $k_{\sm{ion}}$ is the collisional-ionization coefficient 
 \cite{Abelchem}, $f_{\sm{ion}}$ describes ionization from X-rays (see \S \ref{stellar}), and the parametrized case B recombination coefficient for atomic hydrogen $\alpha_H$ is given by
\begin{equation}
\alpha_\fHI=F\times10^{-13}\frac{at^b}{1+ct^d}\ \mathrm{cm}^3\ \mathrm{s}^{-1}
\end{equation} 
with $a=4.309$, $b=-0.6166$, $c=0.6703$, $d=0.5300$ and $t=T/10^4\ K$, which is a fit given in \cite{Pequignot} to the coefficient of \cite{Hummer}. This coefficient takes into account that direct recombination into the ground state does not lead to a net increase of neutral hydrogen atoms, since the photon emitted in the recombination process can ionize other hydrogen atoms in the neighbourhood.  The fudge factor $F=1.14$ serves to speed up recombination and is determined from comparison with the multilevel-code. We further introduce the clumping factor $C(z)\equiv\langle n_e^2\rangle/\langle n_e\rangle^2$ to take into account the increase in the recombination rate in structures of increased density at low redshifts. We use the fit formula
\begin{equation}
 C(z)=27.466\mathrm{exp}(-0.114z+0.001328z^2)\label{clump}
\end{equation}
obtained from simulations of \cite{Mellema} at redshifts $z<40$ and set $C(z)=1$ at higher redshifts. 

The photoionization coefficient $\beta_\fHI$ is calculated from detailed balance at high redshifts as described in \cite{Seager, SeagerFast}. At lower redshifts, however, radiative recombination is no longer balanced by photoionization in the presence of additional energy injection mechanisms like ambipolar diffusion heating. Once the ionized fraction drops below $98\%$, we thus calculate the photoionization coefficient directly from the photoionization cross section given in \cite{Sasaki}. The frequency  $\nu_{\fHI, 2p}$ corresponds to the Lyman-$\alpha$ transition from the $2p$ state to the $1s$ state of the hydrogen atom. The two-photon transition between the states $2s$ and $1s$ is close to Lyman-$\alpha$.  Consequently we use the same frequency for both processes. 
Finally, $\Lambda_\fHI=8.22458\ {\rm s}^{-1}$ is the two-photon rate for the transition $2s$-$1s$ according to \cite{Goldman} and $K_\fHI\equiv \lambda_{\fHI, 2p}^3/[8\pi H(z)]$ the cosmological redshifting of Lyman $\alpha$ photons. The additional terms for Eq. (\ref{temp}) and (\ref{ion}) that are needed to describe the effects of dark matter annihilation and decay will be discussed in \S \ref{darkness}.

The fully-ionized component in the HII regions is described with the volume filling factor $Q_{\fHII}$ that denotes the volume fraction of ionized hydrogen bubbles. It is needed to compute the reionization optical depth and takes UV feedback into account. It evolves as
\begin{equation}
 \frac{dQ_{\fHII}}{dz}=\frac{Q_{\fHII}C(z)n_{\fe,\fHII}\alpha_A}{H(z)(1+z)}+\frac{dn_{ph}/dz}{n_{\fHI}}\label{volumefilling}
\end{equation}
\citep{Shapiro, Haiman, Barkana, Loeb, Choudhury, Schneider}. The UV photon production rate $dn_{ph}/dz$ will be described in \S \ref{stellar} in more detail, $n_{\fe,\fHII}$ denotes the mean electron number density in fully-ionized regions, $n_\fHI$ the neutral hydrogen density in regions that are still unaffected from UV photons and $\alpha_A=4.2\times10^{-13}[T_{max}/10^4\ \mathrm{K}]^{-0.7}\ \mathrm{cm^3/s}$ is the case A recombination coefficient \citep{Osterbrock} which we have chosen here, as recombinations will preferably occur in dense regions where the recombination photons are unlikely to escape into the IGM \citep{Miralda}. It is evaluated at the temperature with $T_{max}=\mathrm{max}(10^4\ \mathrm{K},T)$ to account for the fact that the ionized regions should be heated at least to $10^4$ K. If heating via ambipolar diffusion and decaying MHD turbulence increased the gas temperature above this threshold, it is evaluated at the gas temperature instead.
To compare the models with the observational constraints from WMAP, we calculate the Thomson scattering optical depth of free electrons, given as
\begin{equation}
 \tau_T=\frac{n_{H,0}c}{H_0}\int_0^{z_{\sm{re}}} x_{eff}\sigma_T\frac{(1+z)^2}{\sqrt{\Omega_\Lambda+\Omega_m(1+z)^3}}dz,\label{taure}
\end{equation}
where $n_{H,0}$ is the comoving number density of ionized and neutral hydrogen, $\Omega_\Lambda$ and $\Omega_m$ the usual cosmological density parameters, $H_0$ the Hubble constant and $z_{\sm{re}}$ is the redshift where reionization starts, which we define as the point where the effective ionized fraction $x_{eff}=Q_{\fHII}+(1-Q_{\fHII})x_\fpp$ becomes larger than $6.5\%$. This point has been chosen after a comparison of the actual TE cross calibration spectra with those assuming sudden reionization, but the results are not very sensitive to this choice. In summary, we adopt the following picture:
\begin{itemize}
\item The partially-ionized gas unaffected by UV feedback is modeled with Eqs (\ref{temp}), (\ref{ion}) that describe the evolution of its temperature and ionization degree.
\item The fully-ionized gas is described by the volume-filling factor $Q_{\fHII}$ calculated from Eq. (\ref{volumefilling}). Its temperature is given as the maximum of $10^4$~K and the temperature of the partially-ionized gas.
\item Recombination both in the partially- and fully-ionized gas is enhanced by the clumping factor given in Eq. (\ref{clump}). 
\item For the partially-ionized component, recombination is described using the case B recombination coefficient, appropriate for low-density gas far from the ionizing sources.
\item For the fully-ionized gas, we use the case A recombination coefficient, assuming that most recombinations take place in high-density regions near the ionizing sources.
\end{itemize}
The implementation of different feedback mechanisms is described below in more detail.

\subsection{The generalized filtering mass}\label{filter}
The universe becomes reionized due to stellar feedback. We assume here that the star formation rate (SFR) is proportional to the change in the fraction of collapsed halos $f_{\sm{coll}}$. As shown in \cite{Press}, $f_{\sm{coll}}$ is given as
\begin{equation}
f_{\sm{coll}}=\rm{erfc}\left[\frac{\delta_c(z)}{\sqrt{2}\sigma(m_{\sm{min}})} \right],\label{fcoll}
\end{equation}
where
\begin{equation}
\sigma^2(m)=\int_0^\infty\frac{dk}{2\pi^2}k^2P_{lin}(k)\left[\frac{3j_1(kR)}{kR} \right]^2\;,\label{sigma}
\end{equation}
and where $m_{\sm{min}}$ is the minimum mass of haloes that are able to collapse at a given redshift. Here, $j_1(x)=(\sin x-x\cos x)/x^2$ is the spherical Bessel function, $R$ is related to the halo mass $M_h$ via $M_h=4\pi\rho R^3/3$, $\rho$ is the mean density, $\delta_c=1.69/D(z)$ the linearized density threshold for collapse in the spherical top-hat model and $D(z)$ the linear growth factor. In the absence of magnetic field, $m_{\sm{min}}$ is determined by the filtering mass \citep{GnedinHui, Gnedin}. As discussed in the introduction, tangled magnetic fields can potentially create more small scale structure via the Lorentz force. The calculations by \cite{Tashiro} show that this effect is pronounced in the minihalo regime. While they adopted a constant minimal collapse mass of $10^6 h^{-1}\ M_\odot$ which is independent of the magnetic field, we use the framework of \cite{Gnedin} to take into account the change in the mass scale of halos that can form stars. Indeed, we find that the mass scale is changed significantly (see Fig. \ref{field}), such that the additional small-scale structure from tangled magnetic fields does not contribute to star formation. We introduce the magnetic Jeans mass, \citep{Sethi05}
\begin{equation}
M_J^B\sim10^{10}M_\odot\left(\frac{B_0}{3\times10^{-9}\ \mathrm{G}} \right)^3,
\end{equation}
and the thermal Jeans mass, 
\begin{equation}
M_J=2M_\odot \left(\frac{c_s}{0.2\ \mathrm{km/s}} \right)^3\left(\frac{n}{10^3\ \mathrm{cm}^{-3}} \right)^{-1/2}.
\end{equation}
The filtering mass in the presence of magnetic fields is then given as
\begin{equation}
M_{F,B}^{2/3}=\frac{3}{a}\int_0^a da' M_g^{2/3}(a')\left[1-\left(\frac{a'}{a} \right)^{1/2} \right],
\end{equation}
where $a=1/(1+z)$ is the scale factor and $M_g={\rm{max}}(M_J,M_J^B)$. The minimum halo mass to consider is given as $m_{\sm{min}}={\rm{max}}(M_{F,B},m_{\sm{cool}})$, where $m_{\sm{cool}}$ denotes the minimal halo mass for which the baryons can efficiently cool after collapse. We adopt here the fiducial value $m_{\sm{cool}}=10^5\ M_\odot$ \citep{Greif}. Similar estimates have been given in \cite{Yoshida}, \cite{Barkana} and \cite{Mackey}. For the subsequent analysis, we furthermore assume that only a certain fraction $f_*$ of the collapsing halo mass turns into stars. 

\subsection{Stellar feedback}\label{stellar}
As X-ray photons have long mean free paths, they can play an important role in the ionization and heating of the gas. Assuming that the local correlation between the SFR and the X-ray luminosity (from $0.2-10$ keV) holds up to a renormalization factor $f_X$ \citep[see][]{Furlanetto, Grimm, Ranalli, Gilfanov, Glover}, the X-ray heating function $L_X$ is given as
\begin{equation}
 L_X=3.4\times10^{40}f_X\left(\frac{{\rm{SFR}}}{1\ M_\odot\ \mathrm{yr}^{-1}} \right)\ \mathrm{erg}\ \mathrm{s}^{-1}.
\end{equation}
X-ray emission has two major sources, inverse-Compton scattering of CMB photons with relativistic electrons accelerated in supernovae, and high-mass X-ray binaries. The former may play an increasingly important role at high redshifts, as the CMB photons are more energetic at high redshifts \citep{Oh}. 
In the early universe the factor $f_X \sim 0.5$ if $10^{51}$ erg are released per $100\ M_\odot$, corresponding to an overall efficiency of 5\% \citep{Koyama}. The abundance of X-ray binaries depends on metallicity and the stellar initial mass function, and could be especially large if very massive Pop. III stars dominate \citep{Glover}. In the model presented here, we adopt $f_X\sim0.5$ as a generic value and assume that the uncertainty can be ascribed to the star formation efficiency $f_*$. In general, X-rays lose their energy through three channels. A fraction $f_{X,h}$ goes into heating, a fraction $f_{\sm{X,ion}}$ into ionization, and a fraction $f_{\sm{X,coll}}$ into excitation. These parameters should not be confused with $f_{\sm{ion}}$ introduced in Eq. (\ref{ion}) and $f_{\sm{coll}}$ introduced in Eq. (\ref{fcoll}). We calculate them using the fit formulae in \cite{ShullSteen}. We can thus write
\begin{equation}
 \frac{2 L_X}{3 k_B n_{\rm H}}=10^3\ \mathrm{K}\; f_X\left(\frac{f_*}{0.1}\frac{f_{X,h}}{0.2}\frac{df_{\sm{coll}}/dz}{0.01}\frac{1+z}{10}  \right)H(z).
\end{equation}
The contribution from the X-rays to the term $f_{\sm{ion}}$ in Eq. (\ref{ion}) is then given as
\begin{equation}
f_{\sm{ion}}\sim\frac{f_{X,ion}}{f_{X,h}}\frac{L_X}{13.6\ \mathrm{eV}\ H(z)(1+z)}.
\end{equation}
UV photons from stellar sources are in general absorbed locally, only a fraction $f_{\sm{esc}}$ that manages to escape from the first galaxies can contribute to the ionization of the IGM. Again, the production of ionizing photons is associated with the star formation rate \citep{Furlanetto}, yielding a contribution
\begin{equation}
\frac{dn_{ph}/dz}{n_H}\sim\xi \frac{df_{\sm{coll}}}{dz},
\end{equation}
where 
\begin{equation}
\xi=A_{\fHeI}f_*f_{\sm{esc}}N_{\sm{ion}},\label{xi}
\end{equation}
with $A_{\fHeI}=4/(4-3Y_p)=1.22$ and $N_{\sm{ion}}$ is the number of ionizing photons per stellar baryon. 

\subsection{Models for the stellar population}
The nature of the first stellar sources is still under discussion, and even the question whether the primordial IMF is top-heavy or closer to the locally measured IMF is not solved. While for instance \cite{Abel} and \cite{Bromm} found a top-heavy IMF using adaptive-mesh refinement (AMR) or smoothed-particle hydrodynamics (SPH) simulations, it was shown in \cite{Clark} that primordial and low-metallicity gas can fragment if the evolution of the gas is followed further after the formation of the first clump, due to a dip in the equation of state. Based on similar arguments, \cite{Omukai} argued that even a small metallicity fraction can lead to fragmentation in the first protogalaxies. It was further suggested that magnetic fields can have a crucial influence on the primordial IMF \citep{Silk}. Given these uncertainties, we discuss different models for the stellar population. Dark stars have also been suggested as some of the first luminous sources. We will discuss this possibility in more detail in \S \ref{dark star}.

%

\begin{table}[htdp]
\begin{center}
\begin{tabular}{cccl}
model & population & $N_{\sm{ion}}$  & $f_{\sm{esc}}$ \\
\hline
A &  III & 40,000 & 1\\
B &  III & 40,000 & 1 \hfill ($T_{\sm{vir}} < 10^4\,$K)\\
& & & 0.1 \hfill ($T_{\sm{vir}} \ge 10^4\,$K)\\
C &  III/II & 10,000 & 1 \hfill ($T_{\sm{vir}} < 10^4\,$K)\\
& & & 0.1 \hfill ($T_{\sm{vir}} \ge 10^4\,$K)\\
D &  II  & 4,000 & 0.06\\
\hline
\end{tabular}
\end{center}
\caption{Summary of adopted stellar models. The first column gives the model name. The second column indicates the stellar populations that contribute to reionization. The third column gives the number of ionizing photons per baryon used in Eq (\ref{xi}). The fourth column lists  the adopted escape fractions. Model A is a highly extreme case in which we assume all ionizing photons come from massive Pop III stars and can escape the star-forming halo. In model B photons can escape efficiently only from halos less massive than a corresponding virial temperature of $10^4\,$K, while for higher-mass halos, this fraction is reduced to $10\%$. In model C we assume that Pop III as well as Pop II stars contribute to reionization, and we adopt the same escape probabilities as in B. Model D is another extreme case which assumes that all ionizing photons come from low-mass Pop II stars, with escape fraction $6\%$.}
\label{tab:models}
\end{table}%

In models A and B, we assume that reionization is solely due to metal-free massive Pop III stars. This situation can only be examined from a theoretical point of view. Most investigations of UV feedback from high-mass zero-metallicity stars indicate that ionizing photons can easily escape the star-forming halo and drive  large H{\sc II} regions into the IGM. This suggests that the escape fraction in high-redshift galaxies could be very high, of order unity, being much higher than in the present-day universe \citep[for instance][]{Dove, Ciardi, Fujita}. Detailed numerical simulations of \cite{Whalen} show that indeed the shock bounding HII regions of massive Pop. III stars can easily photo-evaporate the minihalo and lead to an escape fraction of one. However, the situation is not fully clear. There are other studies that suggest that the escape fraction could remain small \citep{Wood}, depending on the mass of the collapsing halo. This is why we adopt two different models. Model A is an extreme case, in which we assume that the escape fraction is $100\%$ independent of the halo mass. The more realistic case is probably model B, where we distinguish whether the virial temperature corresponding to the generalized filtering mass is smaller or larger than $10^4$ K. If it is smaller, we still assume that the halo is easily photoevaporated. If it is larger, we assume that most of the mass is kept within the halo and adopt an escape fraction  $f_{\sm{esc}}=0.1$. The virial mass corresponding to $10^4$ K is given as $M_c=5\times10^7M_\odot \left(\frac{10}{1+z} \right)^{3/2}$ \citep{Oh2,Greif}. For model A and B, we adopt a total number of ionizing photons per baryon of $N_{\sm{ion}}=40,000$, following \cite{Bromm2}.

The heavy elements produced by the very first stars will gradually enrich the IGM. It is very likely that some contribution to reionization comes from low-metallicity Pop II stars as well. As these stars are expected to have lower masses than Pop III stars, we introduce model C, which  corresponds to a stellar population of intermediate mass. It has $N_{\sm{ion}}=10,000$ and an escape fraction according to model B, suggesting that such stars may photo-evaporate only low-mass halos. 

Finally, we consider in model D the extreme case of very rapid chemical enrichment and assume cosmic reionization is entirely driven by low-mass Pop II stars. We assume a stellar mass distribution similar to the local  IMF in the Galactic halo  \citep{Scalo,Kroupa,Chabrier} with a metallicity of $1/20$ of the solar value. This corresponds to $N_{\sm{ion}}\sim4,000$. We furthermore adopt escape fractions that are typical for local star forming galaxies. Many upper limits and a few detections have been found observationally \citep{Hurwitz, Heckman, Deharveng, Bland1, Bland2}, suggesting $f_{\sm{esc}}\sim0.06$. A detection from Lyman-break galaxies at $z\sim3$ implied a higher fraction of $10\%$ \citep{Steidel}, while more recent observations place upper limits of $5-10\%$ or claim detections at even lower levels \citep{Giallongo, Malkan, Fernandez, Inoue}. As the relevant rates scale only linear in $f_{\sm{esc}}$, it seems reasonable to adopt $f_{\sm{esc}}=0.06$ for our Pop II model. 

As it is unlikely that cosmic reionization occurs instantaneously with the onset of Pop III star formation nor that chemical enrichment is so rapid that all ionizing photons come from metal-enriched Pop II stars, we adopt the intermediate scenario C with a mixed population as our fiducial model. All global cosmological parameters are chosen according to the WMAP 5 year data \citep{Komatsu}.

\section{The effect of magnetic fields on the IGM}\label{bfields}
As discussed in \cite{Sethi05,Sethi08}, magnetic fields can significantly alter the thermal evolution of the IGM via ambipolar diffusion heating and decaying MHD turbulence, which can have a significant influence on the filtering mass scale. We explain our treatment of these heating terms in this section.
\subsection{Ambipolar diffusion heating}\label{ambi}
The presence of magnetic fields introduces two different contributions to the heating rate, one coming from ambipolar diffusion and one resulting from the decay of MHD turbulence.\\ In the first case, the contribution can be calculated as \citep{Sethi05, Cowling}:
\begin{equation}
L_{\sm{ambi}}=\frac{\rho_n}{16\pi^2\gamma \rho_b^2 \rho_i}\left | \left(\nabla \times \vec{B} \right)\times \vec{B} \right |^2.\label{ambiheat}
\end{equation}
Here, $\rho_n$, $\rho_i$ and $\rho_b$ are the mass densities of neutral hydrogen, ionized hydrogen and all baryons. The ion-neutral coupling coefficient for primordial gas is given as \citep{Draine, Shang}
\begin{eqnarray}
\gamma=\frac{\frac{1}{2}n_H\langle \sigma v \rangle_{\fHII,\fHI}+\frac{4}{5}n_\fHeI\langle \sigma v \rangle_{\fHII,\fHeI}}{m_H\left[n_\fHI+4n_\fHeI \right]},
\end{eqnarray}
where $n_\fHeI$ is the number density of \HeI and $m_H$ the mass of the hydrogen atom. Collisions with electrons are neglected here, as their contribution is suppressed by a factor $m_e/m_H$. We adopt the zero drift velocity momentum transfer coefficients of \cite{Pinto} for collisions of \HII with \HI and \HeI, which is a good approximation in the absence of shocks. They are given by
\begin{eqnarray}
\langle \sigma v \rangle_{\fHII,\fHI} &=& 0.649T^{0.375}\times10^{-9}\ \mathrm{cm}^3\ \mathrm{s}^{-1},\\
\langle \sigma v \rangle_{\fHII,\fHeI} &=& \big(1.424+7.438\times10^{-6}T\\
&-&6.734\times10^{-9}T^2\big)\times10^{-9}\ \mathrm{cm}^3\ \mathrm{s}^{-1}.\nonumber
\end{eqnarray}
As the power spectrum for the magnetic field is unknown and eq. (\ref{ambiheat}) cannot be solved exactly, we adopt a simple and intuitive approach to estimate the integral for a given average magnetic field $B$ with coherence length $L$. The heating rate can then be evaluated as 
\begin{equation}
L_{\sm{ambi}}\sim\frac{\rho_n}{16\pi^2\gamma \rho_b^2 \rho_i}\frac{B^4}{L^2}. \label{ambiheatapprox}
\end{equation}
The coherence length $L$ is in principle a free parameter that depends on the generation mechanism of the magnetic field. It is constrained through the fact that tangled magnetic fields are strongly damped by radiative viscosity in the pre-recombination universe on scales smaller than the Alfv\'{e}n damping scale $k_{\sm{max}}^{-1}$  given by \citep{Jedamzik, Subramanian, Seshadri}
\begin{eqnarray}
k_{\sm{max}}&\sim&234\ \mathrm{Mpc}^{-1}\left(\frac{B_0}{10^{-9}\ \mathrm{G}} \right)^{-1}\left(\frac{\Omega_m}{0.3}\right)^{1/4}\nonumber\\
&\times&\left(\frac{\Omega_b h^2}{0.02}\right)^{1/2}\left(\frac{h}{0.7} \right)^{1/4},\label{minlength}
\end{eqnarray}
where $B_0=B/(1+z)^2$ denotes the comoving magnetic field. In fact, we expect fluctuations to be present on all scales. As the heating term goes as $L^{-2}$, we thus adopt the minimal value $L=k_{\sm{max}}^{-1}/(1+z)$.

\subsection{Decaying MHD turbulence}\label{decay}
For decaying MHD turbulence, we adopt the prescription of \cite{Sethi05}, 
\begin{equation}
L_{\sm{decay}}=\frac{B_0(t)^2}{8\pi}\frac{3\tilde{\alpha}}{2}\frac{[\ln(1+t_d/t_i)]^{\tilde{\alpha}} H(t)}{[\ln(1+t_d/t_i)+\ln(t/t_i) ]^{\tilde{\alpha}+1}},\label{decayheat}
\end{equation}
where $t$ is the cosmological time at redshift $z$, $t_d$ is the dynamical timescale, $t_i$ the time where decay starts, i. e. after the recombination epoch when velocity perturbations are no longer damped by the large radiative viscosity, $z_i$ is the corresponding redshift. For a power spectrum of the magnetic field strength with power-law index $\alpha$, the parameter $\tilde{\alpha}$ is given as $\tilde{\alpha}=2(\alpha+3)/(\alpha+5)$ \citep{Olesen, Shiromizu, Christensson, Banerjee03}. In the generic case, we expect the power spectrum of the magnetic field to have a maximum at the scale of the coherence length, and the heat input by MHD decay should be determined from the positive slope corresponding to larger scales \citep{Mueller, Christensson, Banerjee03, Banerjee04}. We thus adopt $\alpha=3$ for the calculation. We estimate the dynamical timescale as $t_d=L/v_A$, where $v_A=B/\sqrt{4\pi \rho_b}$ is the Alv\'{e}n velocity and $\rho_b$ the baryon mass density. The evolution of the magnetic field as a function of redshift can be determined from the magnetic field energy $E_B=B^2/8\pi$, which evolves as \citep{Sethi05}
\begin{equation}
\frac{dE_B}{dt}=-4H(t)E_B-L_{\sm{ambi}}-L_{\sm{decay}}.\label{bfield}
\end{equation}

\subsection{The evolution of the IGM}

\begin{figure}
  \centering
   \subfigure[]{\includegraphics[scale=0.4]{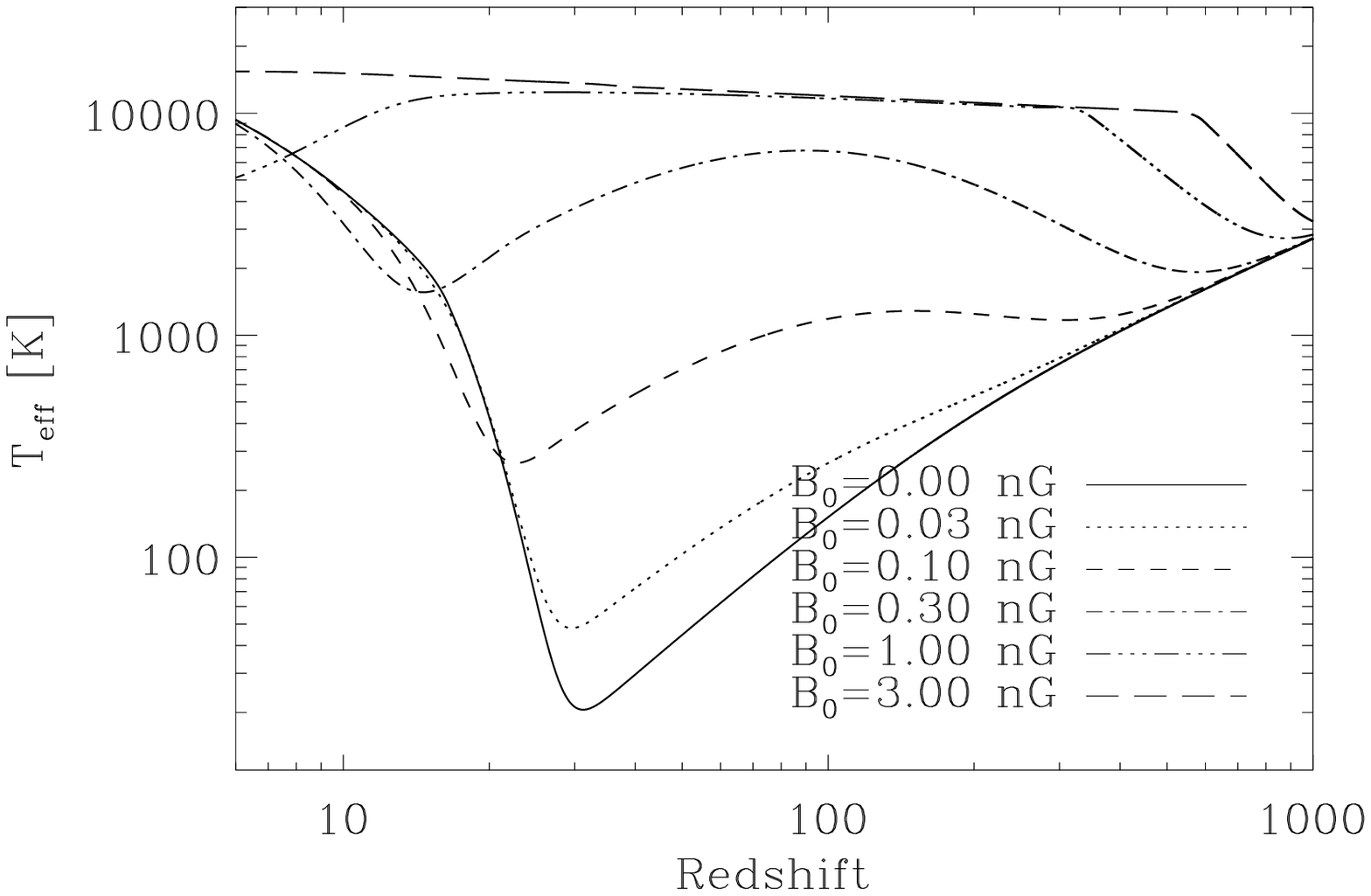}\label{temperature}}\qquad
   \subfigure[]{\includegraphics[scale=0.4]{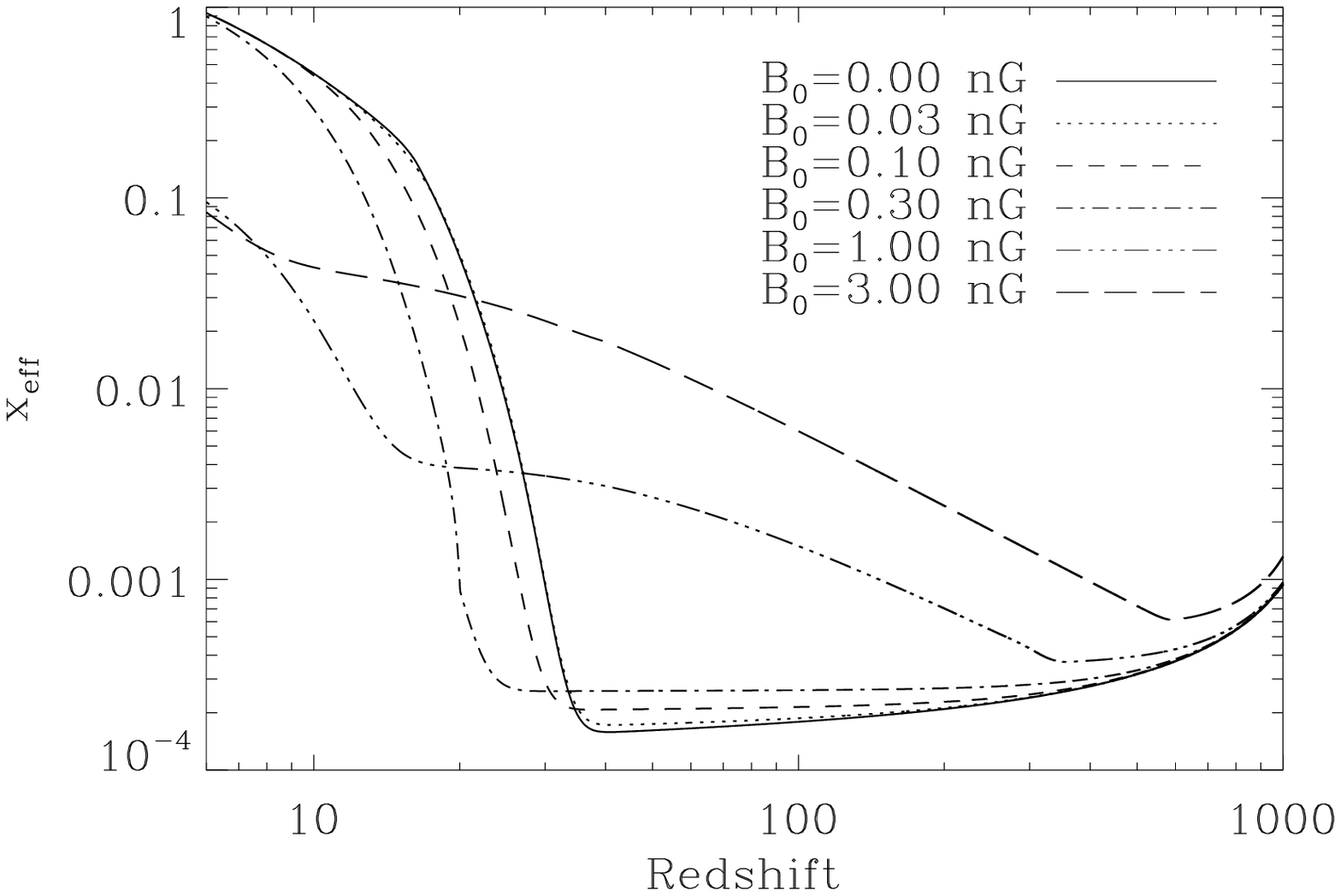}\label{ionization}}\qquad
   \subfigure[]{\includegraphics[scale=0.4]{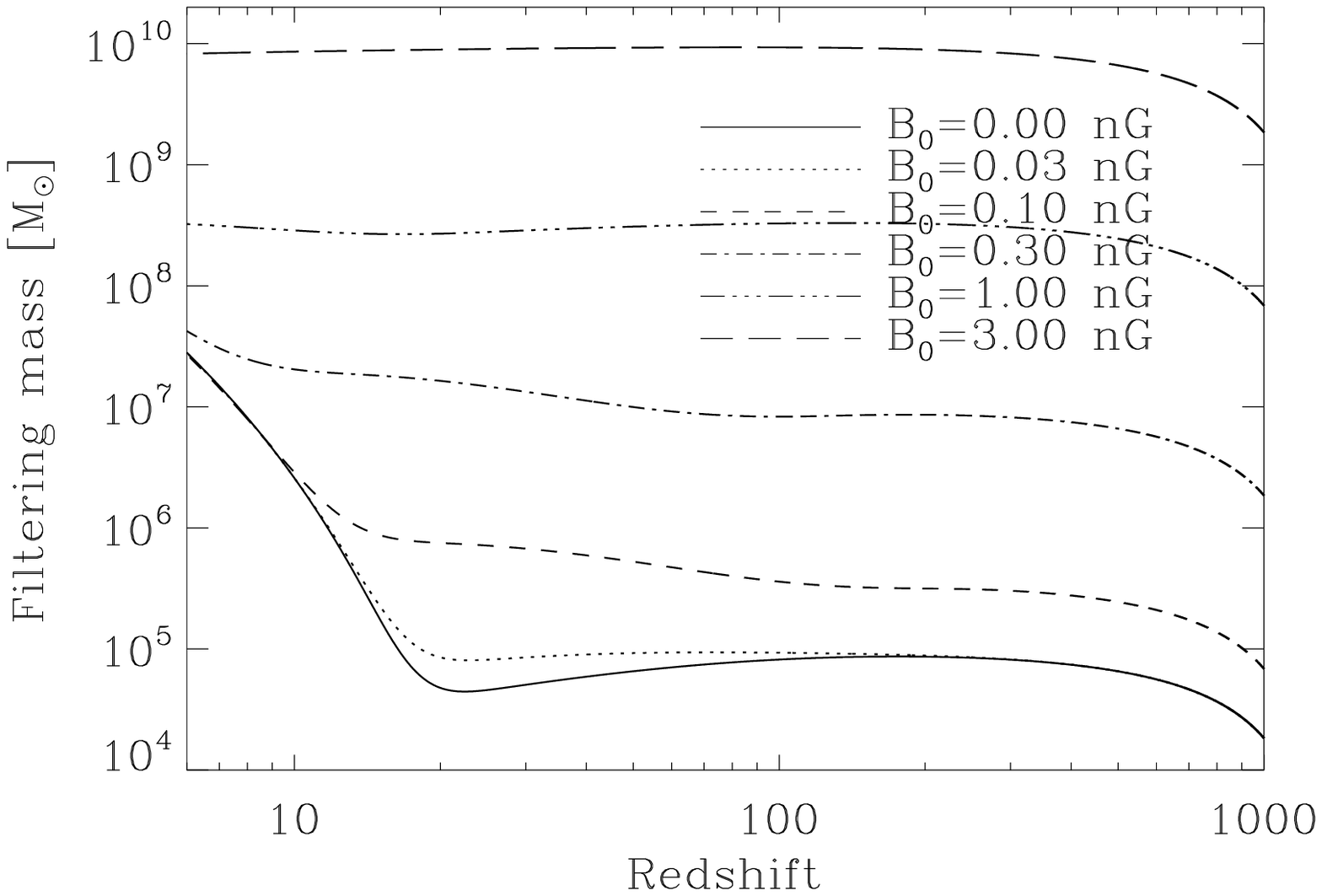}\label{filt}}
\subfigure[]{\includegraphics[scale=0.4]{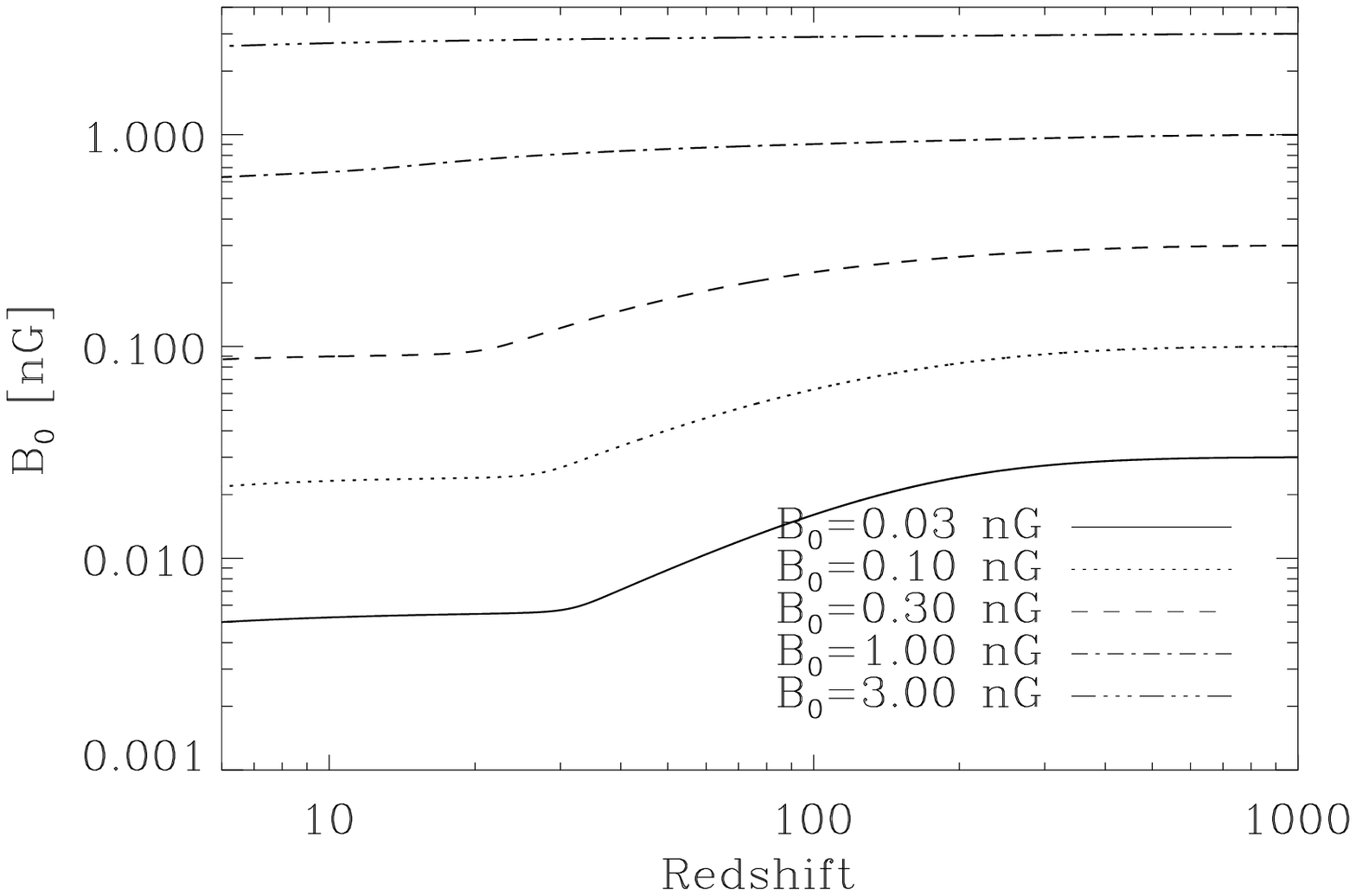}\label{field}}
  \caption{Evolution of different quantities as a function of redshift. a) The effective gas temperature. b) The effective ionized fraction. c) The filtering mass. d) The comoving magnetic field strength.} 
\end{figure}

While the redshift of reionization of course depends on the model for the stellar population, the mechanism which delays reionization in the presence of magnetic fields is always the same. We illustrate this for our fiducal model C adopting a star formation efficiency $f_*$ of 1\%, but point out that the general discussion is valid also for the other stellar population models. We use the cosmological parameters of WMAP 5 \citep{Komatsu}. Fig. \ref{temperature} shows the thermal evolution of the IGM for different magnetic field strengths. In the absence of magnetic fields, the gas temperature follows the temperature of radiation until $z\sim 200$, where Compton scattering becomes inefficient. Afterwards, the gas cools adiabatically until it is reheated during reionization. In the presence of magnetic fields, gas and radiation decouple earlier due to the ambipolar diffusion heating and decaying MHD turbulence and stays at higher temperatures, which prevents collapse in low-mass halos and thus delays reionization. The additional heat increases the ionized fraction at early times (see Fig. \ref{ionization}), while the redshift where the IGM becomes fully ionized is delayed, due to the increased generalized filtering mass (see Fig. \ref{filt}). Collisional ionization introduces a natural temperature plateau of the order $\sim10000$ K, as any further temperature increase will have a backreaction on the ionized fraction of the gas, and thus make ambipolar diffusion less efficient.

\section{Implications from the dark sector}\label{darkness}
A large variety of different particle physics models has been suggested to explain the dark matter content of the universe 
\citep[][]{Olive}. Many of these models predict some interactions in the dark sector and include some form of dark matter decay and annihilation. The consequences of such scenarios for the thermal evolution have been discussed in detail for instance in \cite{Ripamonti, FurlanettoDM}. In this work, we discuss the implications of such scenarios on reionization. As we have seen for the case of primordial magnetic fields, the additional heat input increases the filtering mass and thus increases the minimal halo mass in which the first luminous objects can form, making stellar reionization less effective. On the other hand, once heating through dark matter decay or annihilation is effective, secondary ionization will likely be effective as well and thus increase the Thomson scattering optical depth. As a consequence of dark matter annihilation models, a new phase of stellar evolution was suggested in \cite{Spolyar, Iocco}, in which stars are not powered by nuclear fusion, but by annihilating dark matter within them. These stars could reach masses of up to $10^3\ M_\odot$ \citep{Freese}. Once dark matter annihilation becomes ineffective and such stars enter the main sequence phase, they would thus be extremely bright sources of UV photons. The consequences of such a phase are discussed below. Further models taking into account a dark matter capturing phase have been suggested as well \citep{IoccoBressan, Yoon, FreeseSpolyar}. As they are highly parameter-dependent, we study them in a separate work \citep{SchleicherDark}.
\subsection{Dark Matter annihilation}
The energy of the annihilation products can be deposited into heating, ionization and collisional excitation. The latter is quickly radiated away, but may contribute to the build-up of a Lyman $\alpha$ background that helps to couple the spin temperature of hydrogen to the gas temperature, which may be relevant for $21$ cm observations \citep{FurlanettoDM}. For given models of dark matter, it is possible to work out the detailed absorbed fractions as a function of redshift \citep{Ripamonti}. In this work, however, we adopt the more generic approach of \cite{Furlanetto}. For definiteness, they assume that the dark matter particles annihilate to high-energy photons, which allows to determine the energy fractions going into heat and ionization by the fitting formulae of \cite{ShullSteen}. As they show, heating by dark matter annihilation is accompanied also by a significant increase in the electron fraction due to secondary ionization. For other models of dark matter annihilation, the results can be rescaled by appropriate absorption efficiencies. To Eq. (\ref{temp}) describing the temperature evolution, we thus add a term
\begin{equation}
 \frac{\delta T}{\delta z}=-\frac{2}{3}\frac{\eta_2 m_p c^2}{\eta_1 k_B(1+z)H(z)}\xi_X \chi_h
\end{equation}
according to \cite{FurlanettoDM}, where $\eta_1=1+f_{\fHeI}+x_i$, $\eta_2=1+4f_{\fHeI}$, $f_{\fHeI}$ is the helium fraction by number, $\chi_h$ is the fraction of energy going into heating \citep[see][]{ShullSteen} and $\xi_X$ the effective baryon-normalized ``lifetime'', given as
\begin{equation}
 \xi_X=\frac{\Omega_{DM}\rho_c^0}{m_{DM}}\langle\sigma v\rangle(1+z)^3\left(\frac{\Omega_{DM}}{\Omega_b} \right).
\end{equation}
Here, $\Omega_{DM}$ and $\Omega_b$ denote the cosmological parameters for dark and baryonic matter, $\rho_c^0$ is the critical density at redshift zero, $m_{DM}$ is the dark matter particle mass and $\langle\sigma v\rangle$ the velocity-averaged cross section. In the same way, we add a term to Eq. (\ref{ion}) that describes the evolution of the ionized fraction:
\begin{equation}
 \frac{\delta x_i}{\delta z}=-\eta_2\left(\frac{m_p c^2}{E_{ion}} \right)\xi_X \chi_i,
\end{equation}
where $E_{ion}=13.6$ eV is the hydrogen ionization threshold and $\chi_i$ the fraction of the energy going into ionization, for which we use the fitting formulae of \cite{ShullSteen}. As the decay rate scales with the dark matter density squared, it is most efficient at early times and may thus modify the recombination history, which allows to place upper limits on the dark matter annihilation. As shown in \cite{ZhangAnn}, the WMAP 1-year-data yield an upper limit of
\begin{equation}
 \langle\sigma v\rangle\leq 2.2\times10^{-29}\ \mathrm{cm^3}\ \mathrm{s}^{-1}f_{abs}^{-1}\left(\frac{m_{DM}}{MeV} \right),
\end{equation}
where $f_{abs}$ corresponds to the energy fraction actually absorbed into the IGM and is of order 1 for particle masses in the MeV range \citep{Ripamonti}.

\subsection{Dark Matter decay}
In a similar way, dark matter decay can alter the thermal and ionization history of the universe. The lifetimes of dark matter particles considered here will be considerably larger than the age of the universe to ensure that the abundance of dark matter does not change significantly. Still, the conversion of particle mass to thermal energy can have an important impact on the thermal evolution of the universe \citep{Ripamonti, FurlanettoDM}. In this limit, the decay rate is constant over time and the effect at lower redshifts is more pronounced compared to the case of dark matter annihilation. To calculate the heating and ionization rate, the same formalism can be employed as for dark matter annihilation, but we adopt the baryon-normalized decay rate
\begin{equation}
 \xi_X=\frac{\Omega_{DM}}{\Omega_b t_X},
\end{equation}
where $t_X$ is the lifetime of the dark matter particle. Based on the modified recombination and reionization histories in the presence of dark matter decay, it was shown that \cite{ZhangDec}
\begin{equation}
 \frac{f_\chi f_{abs} }{t_X}\leq 2.4\times10^{-25}\ \mathrm{s}^{-1},
\end{equation}
where $f_\chi$ is the fraction of the particle mass-energy released through decay and $f_{abs}$ the fraction of the released energy deposited into the IGM. If data of large-scale surveys are included, the constraint can be slightly improved, yielding
\begin{equation}
 \frac{f_\chi f_{abs}}{t_X}\leq 1.7\times10^{-25}\ \mathrm{s}^{-1}.
\end{equation}

\subsection{Dark stars}\label{dark star}
Dark stars have been suggested as a possible consequence of dark matter annihilation in the early universe \citep{Spolyar, Iocco, Freese}. If indeed the first stars form in the central peaks of the dark matter distribution in the early minihaloes, they might be powered by dark matter annihilation instead of nuclear fusion. Such stars could have up to $10^3\ M_\odot$ and surface temperatures in the range of $3000-10000$ K, i. e. significantly colder than conventional Pop. III stars. However, this dark phase will gradually come to an end as the dark matter in the cusp annihilates, so the star may contract and finally reach a main sequence phase. A linear stability analysis and 1D simulations including hydrodynamics and radiation indicates the stability of such objects \citep{Gamgami}. In this case, one can expect a similar number of photons per stellar baryon as for conventional Pop. III stars, perhaps even higher by a factor of 2 \citep{Bromm2}. As it is crucial for these objects to form on the peak of the dark matter density, we expect them to form only in low-mass halos with a virial temperature below $10^4$ K, as the so-called atomic cooling halos are very turbulent \citep{Greif} and are more likely to form a stellar cluster instead of a single massive star \citep{Clark}. In addition, such massive halos are more likely to have accreted material from previous metal enrichment \citep{Tornatore}, which may cause fragmentation as well \citep{Omukai}. In previously ionized regions, the thermodynamics of collapse are significantly altered and simulations generally find lower-mass stars \citep{Yoshida1, Yoshida2}, which might be an important limitation for dark stars as well.  

\section{Constraining the parameter space with WMAP 5}\label{cmbfast}
In this section, we show the results for the different stellar populations and in the context of different additional physics.
\subsection{Stellar reionization with and without primordial magnetic fields}

\begin{figure}
  \centering
  {\includegraphics[scale=0.45]{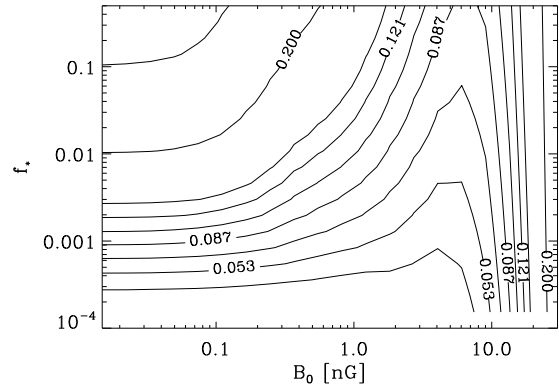}}
\caption{The reionization optical depth for Pop. III stars (model A) in the presence of primordial magnetic fields. The contour lines are equally spaced around $\tau_{\sm{re}}=0.87$ with $\Delta\tau=0.017$, corresponding to the different $\sigma$-errors of the measurement. Magnetic fields higher than $20$ nG are clearly excluded by the optical depth alone. The transition between stellar reionization and collisional ionization occurs at about $5$ nG, providing a more stringent limit if metal enrichment is required.
}\label{tauA}
\end{figure}
\begin{figure}
  \centering
  {\includegraphics[scale=0.45]{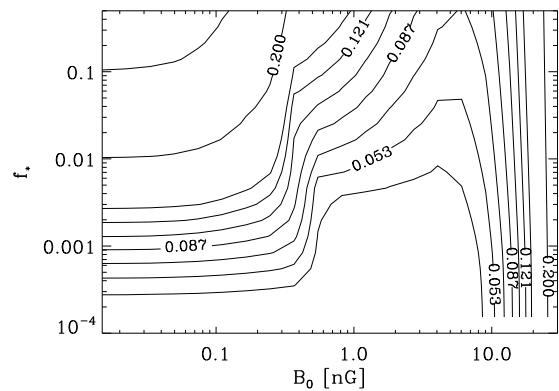}}
\caption{The reionization optical depth for Pop. III stars (model B) in the presence of primordial magnetic fields. Magnetic fields higher than $20$ nG are clearly excluded by the optical depth alone. The transition between stellar reionization and collisional ionization occurs at about $5$ nG. We can further exclude magnetic fields between $2$ and $5$ nG within 3-$\sigma$, as too high star formation efficiencies would be required to obtained the measured optical depth.
%
%
}\label{tauB}
\end{figure}
\begin{figure}
  \centering
  {\includegraphics[scale=0.45]{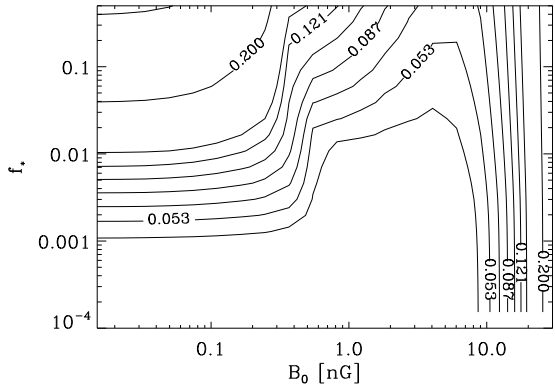}}
\caption{The reionization optical depth for a mixed population (model C) in the presence of primordial magnetic fields. Magnetic fields higher than $20$ nG are clearly excluded by the optical depth alone. The transition between stellar reionization and collisional ionization occurs at about $5$ nG. In addition, magnetic fields between $0.7$ and $5$ nG can be excluded, as they would require unreasonably high star formation efficiencies for the measured optical depth.
}\label{tauC}
\end{figure}
\begin{figure}
  \centering
  {\includegraphics[scale=0.45]{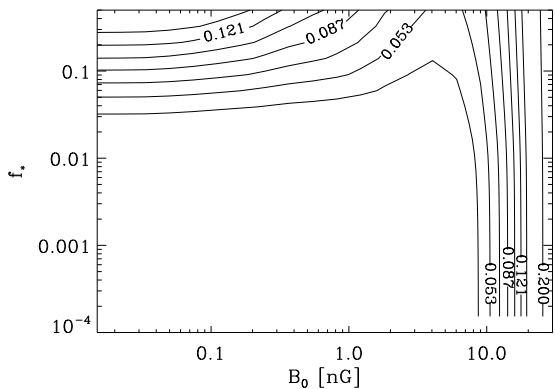}}
\caption{The reionization optical depth for Pop. II stars (model D) in the presence of primordial magnetic fields. For magnetic fields lower than $5$ nG, stellar reionization dominates, but unreasonably high star formation efficiencies would be required to reconcile the measured optical depth. For stronger magnetic fields, collisional reionization dominates, which can be excluded if metal-enrichment is imposed. Magnetic fields larger than $20$ nG can be excluded from the optical depth alone.
}\label{tauD}
\end{figure}

We first concentrate on the effect of primordial magnetic fields and compare the results of different models for the stellar population. 
We run our reionization model for a range of different star formation efficiencies and primordial magnetic fields and obtain the corresponding optical depth. To constrain the parameter space, some assumptions need to be made. The first is obvious: The calculated optical depth must agree with the optical depth measured by WMAP 5 at least within $3\sigma$. In addition, we require that the star formation efficiency may not be unreasonably high.  For Pop. III stars with $\sim100\ M_\odot$ forming in halos of $\sim10^6\ M_\odot$ (dark matter), star formation efficiencies of the order $0.1\%$ are expected. It seems thus reasonable to reject star formation efficiencies higher by one order of magnitude, i. e. of the order $1\%$. Further requirements are that reionization must be complete by redshift 6, and that there should be a stellar contribution to reionization, so that the universe becomes metal-enriched. Now we discuss the constraints resulting from these criteria.\\ \\

For all stellar models, we can exclude magnetic fields larger than $20$ nG, based on the optical depth alone (see Fig. \ref{tauA} to \ref{tauD}). a transition occurs at about $5$ nG: For lower magnetic fields, reionization is still mostly due to stellar radiation, whereas for higher magnetic fields, the IGM is heated to such high temperatures that collisional ionization becomes very efficient and increases the optical depth. In this regime, structure formation is strongly suppressed, as the magnetic Jeans mass scales with $B_0^3$ and dominates over the thermal Jeans mass. Independently of the stellar model, such a regime is always found between $5$ and $20$ nG. Therefore, structure formation is considerably impeded such that metals do not form and the universe is not fully ionized by redshift $6$. We can thus exclude magnetic field strength above $5$ nG independent of the stellar model.\\ \\

 If we further require that the star formation efficiency must be lower than $1\%$, the results become model-dependent. In model A, which is an extreme case where very high escape fractions are assumed even for very massive halos, no further constraint on the magnetic field is possible on the $3\sigma$ level. Still, even in this extreme case, one can reject fields larger than $2.5$ nG on the $1\sigma$ level. The other extreme case, model D with low escape fractions and a low number of ionizing photons per stellar baryon, can be rejected completely, as it always requires unphysically high star formation efficiencies. \\ \\

For more realistic cases, essentially model B and C, we find that a critical magnetic field strength exists above which a very high star formation efficiency is needed to get into the $3\sigma$ interval around the measured optical depth. For model B, the critical value is $2$ nG, for model C, it is at $0.7$ nG. As it seems likely that a transition to less massive Pop. II stars occurs during reionization, model C might indeed be the most realistic case and provide an upper limit of $0.7$ nG.

\subsection{Dark reionization scenarios}
\begin{figure}
  \centering
  {\includegraphics[scale=0.45]{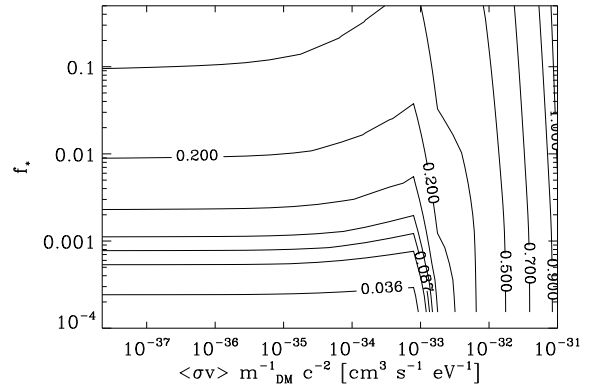}}
\caption{The reionization optical depth for Pop. III stars (model B) in the presence of dark matter annihilation. At $\langle\sigma v \rangle/m_{DM}\sim10^{-33}\ \mathrm{cm}^3/\mathrm{s}/\mathrm{eV}$, secondary ionization of the annihilation products starts to dominate over stellar reionization. Values higher than $3\times10^{-33}\ \mathrm{cm}^3/\mathrm{s}/\mathrm{eV}$ are ruled out by the WMAP 5-year data.
}\label{img:annihilation}
\end{figure}
\begin{figure}
  \centering
  {\includegraphics[scale=0.45]{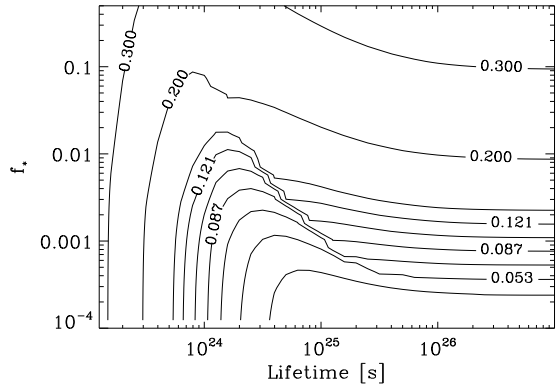}}
\caption{The reionization optical depth for Pop. III stars (model B) in the presence of dark matter decay. For lifetimes lower than $3\times10^{24}$ s, secondary ionization of the annihilation products starts to dominate over stellar reionization. Lifetimes lower than $3\times10^{23}$ s are ruled out by the WMAP 5-year data.
}\label{img:decay}
\end{figure}

For the case of dark matter annihilation and decay, we will show only the results for the Pop. III star model B. Regarding the competition between stellar reionization and secondary ionization from high-energetic annihilation / decay products, this corresponds to a rather conservative choice if one assumes that stellar reionization should do the main contribution. Such a stellar population could be interpreted either as conventional Pop. III stars or as so-called dark stars in a main sequence phase. We find that dark matter annihilation starts to become important for $\langle\sigma v \rangle/m_{DM}\sim10^{-33}\ \mathrm{cm}^3/\mathrm{s}/\mathrm{eV}$ and completely dominates the stellar contribution for higher values (see Fig. \ref{img:annihilation}). Values higher than $3\times10^{-33}\ \mathrm{cm}^3/\mathrm{s}/\mathrm{eV}$ are inconsistent with the WMAP 5-year data. The results for dark matter decay are given in Fig. \ref{img:decay}. Decay starts to become important for lifetimes below $3\times10^{24}$ s.
However, lifetimes smaller than $3\times10^{23}$ s are incompatible with WMAP 5. As pointed out recently, constraints on dark matter annihilation can be significantly improved taking into account the dark matter clumping factor $C(z)=\langle \rho_{\sm{DM}}^2(z) \rangle/\langle \rho_{\sm{DM}}(z)\rangle^2$ \citep{Chuzhoy}. However, this quantity is highly uncertain and may vary by six orders of magnitude \citep{Cumberbatch}. In the framework of light dark matter, we recently showed that is is constrained to be smaller than $10^5$ at redshift zero \citep{SchleicherGlover}, yielding an enhancement of the annihilation rate by a factor of $10$ or more at redshift $20$. Other dark matter models \cite{Cumberbatch} even yield a clumping factor that drops down to $1$ at these redshifts, such that our calculation altogether provides a conservative and firm upper limit.

From Fig. \ref{img:annihilation}, we can further see that very massive dark stars in a main sequence phase, which should have higher star formation efficienes that are an order of magnitude higher than conventional Pop. III stars, are ruled out by WMAP 5, as the models require rather low star formation efficiencies to reproduce the Thomson scattering optical depth. Of course, this conclusion holds only if one assumes that reionization is completely due to dark stars, which seems rather unlikely (see discussion in \S \ref{dark star}), and we will discuss some alternatives in \S \ref{outlook}.

\section{Conclusions and outlook}\label{outlook}
We have calculated the reionization optical depth for different stellar models in the presence of primordial magnetic fields as well as dark matter annihilation and decay. The results indicate which star formation efficiencies are required in the presence of some additional heating mechanism for a given stellar model. Considering different stellar models and primordial magnetic fields, we find the following results:
\begin{enumerate}
\item Independent of the model for the stellar population, we can 
securely 
exclude primordial magnetic fields larger than $5$ nG.
\item For the most realistic case with a mixed stellar population (model C), we 
even
find an upper limit of $0.7$ nG, as higher magnetic fields would require star formation efficiencies larger than $1\%$, which is unrealistic. Similar results are found for model B, assuming reionization completely due to Pop. III stars.
\item Reionization only due to population II stars (model D) is ruled out completely.
\end{enumerate}
For dark matter, we found the following results:
\begin{enumerate}
\item Dark matter annihilation provides noticeable contributions to the reionization optical depth only for thermally averaged mass-weighted cross sections $\langle\sigma v \rangle/m_{DM}\ge10^{-33} \mathrm{cm}^3/\mathrm{s}/\mathrm{eV}$.
\item Parameters $\langle\sigma v \rangle/m_{DM}\ge3\times10^{-33}\ \mathrm{cm}^3/\mathrm{s}/\mathrm{eV}$ can be ruled out by $3\sigma$ on the basis of WMAP 5 year data.
\item Dark matter decay becomes important for the reionization optical depth for lifetimes below $3\times10^{24}$ s.
\item Dark matter lifetimes below $3\times10^{23}$ s are ruled out by $3\sigma$. 
\end{enumerate}
These results are in agreement with conservative constraints obtained from the gamma-ray background \citep{Mack}. We further showed that reionization can not be due to $\sim1000\ M_\odot$ dark stars alone, as the corresponding optical depth would be significantly too high. One might wonder whether heating from dark matter annihilation might help to significantly delay stellar reionization, in order to reconcile this model with observations. However, as can be seen in Fig. \ref{img:annihilation}, this is more than compensated by the effects of secondary ionization, once that dark matter annihilation starts to have a significant influence on the IGM. In case collider experiments like the LHC \footnote{http://lhc.web.cern.ch/lhc/} or other dark matter detection experiments \footnote{A list of dark matter detection experiments is given at http://cdms.physics.ucsb.edu/others/others.html.} find evidence for a self-annihilating dark matter candidate, this can be seen as evidence for a rapid transition towards a different mode of star formation, or a problem in our understanding of dark stars.  To reconcile dark stars with observations, the following scenarios seem feasible:
\begin{enumerate}
\item Dark stars with $1000\ M_\odot$ may be extremely rare objects, and their actual mass scale is closer to the mass scale for typical Pop. III stars. In such a case, reionization could not distinguish between dark stars and conventional Pop. III stars. 
\item The transition to lower-mass stars occurs very rapidly, such that dark stars only contribute to the very early phase of reionization. Reasons for that might be chemical, radiative as well as mechanical feedback (see also the discussion in \S \ref{dark star}). In fact, even a double-reionization scenario might be conceivable, in which very massive dark stars ionize the universe at high redshifts. Due to chemical and radiative feedback, formation of such stars might be suppressed and the universe might become neutral again, until the formation of less massive stars becomes efficient enough to reionize the universe.
\item Dark stars do not reach a main-sequence phase, but are disrupted earlier by some non-linear instability. Such an instability would be constrained by the fact that it neither appears in a linear stability analysis nore in 1D simulations. Violent explosions of dark stars might even be considered as a source for Gamma-Ray bursts.
\end{enumerate}
While too definite conclusions on the existence of dark stars are not yet possible, it is at least indicated that the possibilities mentioned above should be explored in more detailed, and a better understanding of the properties of the dark stars. A better understanding of their evolution after the dark phase will certainly help to better understand their possible role during reionization. 

The constraints derived here are independent of other works that essentially rely on the physics of recombination to derive upper limits on additional physics \citep[e. g.][]{Barrow97, ZhangAnn, ZhangDec}. In particular, magnetic fields can evolve dynamically and their field strength may thus change between these epochs. It has recently been suggested that the Biermann Battery effect creates magnetic fields in the presence of an electron pressure gradient\cite{Xu}. We thus need to probe magnetic fields at different epochs. As we will show in a separate work \citep{Schleicher}, upcoming $21$ cm measurements will allow to probe the thermal history before and during reionization in great detail, and may allow to detect primordial magnetic fields of the order of $0.1$ nG. 

Additional ways of probing the reionization history and the dark ages exist as well: Scattering of CMB-photons in fine-structure lines of heavy elements may lead to a frequency-dependent CMB power spectrum and may allow to measure metal-abundances as a function of redshift \citep{Basu04}. Before reionization, molecules may form in the IGM and introduce further frequency-dependent features in the CMB \citep{Schleicher2}. Such features are likely enhanced in the presence of either primordial magnetic fields or dark matter decay / annihilation, as the increased electron fraction catalyses the formation of molecules, and the additional heat input leads to a departure of the level populations from the radiation temperature.

Further improvements are expected from the upcoming measurement of Planck, that will measure the reionization optical depth with unprecedent accuracy and thus allow to strengthen the constraints obtained here on the stellar populations and additional physics. In addition, a more accurate determination of the cosmological parameters will remove further uncertainties in the present models of reionization.

\begin{acknowledgments}
We thank Paul Clark, Andrea Ferrara, Katherine Freese, Daniele Galli, Simon Glover, Thomas Greif and Francesco Palla for
many exciting discussions on this topic. DRGS thanks the Heidelberg
Graduate School of Fundamental Physics (HGSFP), the LGFG for
financial support and the INAF-Osservatorio Astrofisico di Arcetri for financial support. The HGSFP is funded by the Excellence Initiative of
the German Government (grant number GSC 129/1). RB is funded by the
Emmy-Noether grant (DFG) BA 3607/1. RSK thanks for support from the 
Emmy Noether grant KL 1358/1. All authors also acknowledge subsidies 
from the DFG SFB 439 {\em Galaxies in the Early Universe}. We thank the anonymous referee for help improving the manuscript.
\end{acknowledgments}

\end{document}